\pdfoutput=1
 
 \documentclass[10pt,twocolumn]{article} 
\usepackage{amssymb} 
\usepackage[figuresright]{rotating} 
 
\def\gsim{\ \rlap{\raise 3pt \hbox{$>$}}{\lower 3pt \hbox{$\sim$}}\ }
\def\lsim{\ \rlap{\raise 3pt \hbox{$<$}}{\lower 3pt \hbox{$\sim$}}\ }

\newcommand{\be}{\begin{equation}}
\newcommand{\ee}{\end{equation}}
\newcommand{\bea}{\begin{eqnarray}}
\newcommand{\eea}{\end{eqnarray}}

\newcommand{\eqn}[1]{eq.~(\ref{#1})}
\newcommand{\eqns}[2]{eqs.~(\ref{#1})-(\ref{#2})}

\newcommand{\Eqn}[1]{Eq.~(\ref{#1})}

 \def\chiu{{\raise 2pt \hbox{$\,\chi$}}{\lower 1pt \hbox{$_{u\,}$}}}
 \def\chiuqt{{\raise 2pt \hbox{$\,\chi$}}
            \hbox{$^4\!\!$}{\lower 1pt \hbox{$_{\!u\,}$}}}
\def\chiusq{{\raise 2pt \hbox{$\,\chi^2\!\!$}}{\lower 1pt \hbox{$_{u\,}$}}} 
\def\chid{{\raise 2pt \hbox{$\,\chi$}}{\lower 1pt \hbox{$_{d\,}$}}}
 \def\chidsq{{\raise 2pt \hbox{$\,\chi^2\!\!$}}{\lower 1pt \hbox{$_{d\,}$}}}
 \def\chie{{\raise 2pt \hbox{$\,\chi$}}{\lower 1pt \hbox{$_{e\,}$}}}
 \def\chiesq{{\raise 2pt \hbox{$\,\chi^2\!\!$}}{\lower 1pt \hbox{$_{e\,}$}}}
 \def\chiud{{\raise 2pt \hbox{$\,\chi$}}{\lower 1pt \hbox{$_{u,d\,}$}}}
 \def\chiude{{\raise 2pt \hbox{$\,\chi$}}{\lower 1pt \hbox{$_{u,d,e\,}$}}}

\begin{document}

% \begin{frontmatter} 
 
 
 
%?% \dochead{} 

\title{%\Large\bf
  Quark Yukawa pattern from spontaneous breaking of flavour 
$SU(3)^3$\footnote{Contribution to the proceedings of the $10^{\rm th}$ Simposio
    Latinoamericano de F\'{\i}sica de Altas Energ\'{\i}as (X SILAFAE)
    - Ruta N, Medell\'{\i}n, Colombia, November 24-28 2014.  Work done
    in part with Jose R. Espinosa and Chee Sheng Fong.}
% \tnoteref{collab}
} 
 

% \author[USP]{Chee Sheng Fong} 
% \address[USP]{Instituto de F\'{\i}sica, 
% Universidade de S\~ao Paulo, C.\ P.\ 66.318, 05315-970 S\~ao Paulo, Brazil}
%  \ead{fong@if.usp.br}

\author%[LNF]
{Enrico Nardi \\ [5pt]
% } \address[LNF]{
\it 
INFN, Laboratori Nazionali di Frascati 
  C.P. 13, 100044 Frascati, Italy}
% \tnotetext[collab]{Work done in part with  Jose R. Espinosa 
%  and Chee Sheng Fong.}
%  \ead{enrico.nardi@lnf.infn.it} 

\date{}

\maketitle

\begin{abstract} 
  A $SU(3)_Q \times SU(3)_u \times SU(3)_d$ invariant scalar potential
  breaking spontaneously the quark flavour symmetry can explain the
  standard model flavour puzzle.  The approximate alignment in flavour
  space of the vacuum expectation values of the up and down `Yukawa
  fields' results as a dynamical effect. The observed quark mixing
  angles, the weak CP violating phase, and hierarchical quark masses
  can be all reproduced at the cost of introducing additional
  (auxiliary) scalar multiplets, but without the need of introducing
  hierarchical parameters.
\end{abstract}

% \begin{keyword} 

%%Spontaneous Symmetry Breaking \sep
%%Quark masses and mixing angles \sep
%%Beyond the Standard model. 

 
% \end{keyword} 
 
% \end{frontmatter} 
 
% \linenumbers 
 

\thispagestyle{empty}

\section{Introduction} 
\label{sec:intro} 

Fermion family replication represents probably the oldest unexplained
puzzle in elementary particle physics, dating back to the discovery of
the muon by Anderson and Neddermeyer at Caltech in 1936.  With the
discovery of all the other second and third generation particles, the
puzzle became even more intriguing because fermions with the same
$SU(3)_C \times SU(2)_L\times U(1)_Y$ quantum numbers have been found
with mass values that span up to five orders of magnitude.
% (leaving aside neutrino masses, which we do not discuss in this paper).  
Explaining such strongly hierarchical mass patterns requires a
more fundamental theory than the Standard Model (SM), and  a plethora of
attempts in this direction have been tried. In their
large majority they basically follow two types of approaches:

\begin{enumerate}
\item[(i)] The first is to postulate new symmetries under which
  fermions with the same SM quantum numbers transform differently. The
  fact that fermion families appear to replicate is then just an
  illusory feature of the low energy theory, ascribable to our
  incomplete knowledge of the full set of fundamental quantum numbers.
  This is, for example, the basic ingredient of the popular
  Froggatt-Nielsen mechanism~\cite{Froggatt:1978nt}, in which the
  hierarchy of the Yukawa couplings follows from a dimensional
  hierarchy in the corresponding effective Yukawa operators, obtained
  by assigning to the lighter generations larger values of new Abelian
  charges.

\item[(ii)] A different approach is to assume that the three
  generations contain exact replica of the same states. The gauge
  invariant kinetic term for each type of fermions (same charge and
  chirality) is characterized by a $U(3)$ (flavour)
  symmetry~\cite{Chivukula:1987py}. When this symmetry is broken
  explicitly via the Yukawa terms, we have the SM. A more interesting
  idea is that the flavour symmetry is broken spontaneously (SFSB) by
  vacuum expectation values (vevs) of scalar `Yukawa fields',
  transforming under the various $U(3)$ in such a way that at the
  Lagrangian level, the flavour symmetry is exact.
\end{enumerate} 

The first approach basically relies on {\it ad hoc} assignments of new
quantum numbers in order to reproduce qualitatively the observed mass
patterns.  The second approach can be considered theoretically more
ambitious (as it relies on less {\it ad hoc} assumptions) although it
is by far more challenging than the first one regarding successful
model implementations.  In order to offer a natural solution to the
Yukawa hierarchy, such models should not rely on a hierarchical
arrangement of parameters or on some tuning between them.  Loop-induced
hierarchies for example would be plausible~\cite{Nardi:2011st},
but we have found that this possibility is
vetoed~\cite{Espinosa:2012uu}.  Dynamical mechanisms inducing strong
suppressions of some parameters are another possibility which we have
instead proved being viable~\cite{Fong:2013dnk}.

The idea that quark masses could arise from the minimum of a scalar
potential invariant under a suitable symmetry is in fact rather old,
and group theoretical methods to identify the natural extrema of a
 `Yukawa potential' were established already in the early
seventies~\cite{Michel:1971th,CabibboMaiani:1970}. Nowadays the
literature on attempts towards a dynamical explanation of the Yukawa
couplings (employing different flavour groups and different
flavour-breaking fields) is
extensive~\cite{Anselm:1996jm,Berezhiani:2001mh,%
  Koide:2008qm,Koide:2008tr,%
  Koide:2012fw,Feldmann:2009dc,Albrecht:2010xh,Grinstein:2010ve,%
  Alonso:2011yg,deMedeirosVarzielas:2011wx,Mohapatra:2012km,%
  Alonso:2013nca}.  

The present contribution is based on the three
papers~\cite{Nardi:2011st,Espinosa:2012uu,Fong:2013dnk} and it
describes the main steps (difficulties, vetoes, mandatory
requirements) that lead us to understand the features needed to
implement SFSB in what we think is the best motivated scenarios, which
is based on the quark flavour symmetry:
\begin{equation}
  \label{eq:GF}
\mathcal{G}_{\cal F} = SU(3)_Q\times SU(3)_u\times SU(3)_d  \,, 
\end{equation}
where $Q$, ($u$ and $d$) denote the $SU(2)_L$ quark doublets (singlets).
Quarks couple to the Higgs field via the effective operator 
\begin{eqnarray}
\label{eq:Ly}
-{\cal L}_{Y} & = & 
\sum_{q=u,d} \left[\,\frac{1 %\tilde\kappa_q
}{\Lambda}\, \overline{Q}\,Y_{q}
\,q\, H_q+{\rm h.c.}\,\right],
\end{eqnarray}
where $H_d=H$ is the Higgs field ($H_u=i\sigma_2 H$),
% $\tilde\kappa_{u,d}$ are dimensionless (complex) couplings, 
$Y_{u,d}$ are the up- and down-type Yukawa fields, and $\Lambda$ is
the scale where the effective operators arise.  The theoretical
challenge is now finding a ${\cal G}_{\cal F} $-invariant scalar
potential $V(Y_q,Z)$ (where $Z$ denotes generically additional scalars
coupled to $Y_q$ in a symmetry invariant way) which can break ${\cal
  G}_{\cal F} $ spontaneously and yield a set of vevs $\langle
Y_{q}\rangle$ reproducing the observed structure of the SM Yukawa
couplings.

\section{A single Yukawa field: generating a hierarchy}
\label{sec:singleY}

Let us start by considering a single Yukawa multiplet, e.g. $Y=Y_u$,
that under the relevant flavour symmetry 
\begin{equation}
\label{eq:GLR}
\mathcal{G}_{LR} = SU(3)_L\times SU(3)_R
\end{equation}
transforms as $(3,\bar 3)$.  As a first step, we want to explore the
possibility of generating a Yukawa hierarchy via SFSB of
$\mathcal{G}_{LR}$.

\subsection{Minimization of the tree level potential }
\label{sec:tree}

It is convenient to parameterize $Y$ 
by means of its singular value decomposition:
\begin{eqnarray}
  \label{eq:biunitary}
 Y &=& {\cal V}^\dagger\, \chi\,  {\cal U} \,,\  \\ 
\chi &=& {\rm diag}\left(u_1,\,u_2,\,u_3\right)\,, 
\end{eqnarray}
where the matrices $ {\cal V}$ and $ {\cal U}$ are unitary and the
entries in $\chi$ are real nonnegative.  We can write down three
renormalizable invariants with respect to the $\mathcal{G}_{LR}$
transformations $Y \to V_L\,Y\, V_{R}^\dagger$ (with $\det
V_{L,R} = +1$):% ~\cite{Nardi:2011st}:
\begin{eqnarray}
\label{eq:Tu}
  T &=& {\rm Tr}( YY^\dagger)= \sum_i u_i^2\,, \\ 
\label{eq:Au}
  A &=& {\rm Tr}\left[{\rm Adj}( YY^\dagger)\right]= 
\frac{1}{2}\sum_{i\neq j} u^2_iu^2_j\,,\\  
\label{eq:Du}
{\cal D} &=& {\rm Det}(Y)= e^{i\delta}\,\prod_i u_i 
\equiv e^{i\delta}\,D\,, 
\end{eqnarray}
where $\delta= {\rm Arg}\left[{\rm Det}\left({\cal V}^\dagger{\cal
      U}\right)\right]$ and $D=\left|\mathcal{D}\right|$.  The most
general potential reads:
\begin{equation}
  \label{eq:V3}
%  \hat V &=&  \Lambda^4\,
V = 
% \Lambda^4\left(
\lambda \left[T- \frac{m^2}{2\lambda}\right]^2 + 
 \tilde\lambda_A A + 
% 2\,\mu\,\cos(\phi+\delta)\,D \,.
 \tilde \mu\, {\cal D} +  \tilde \mu^*\, {\cal D}^* \,, 
% \right)
\end{equation}
and by taking $\tilde \mu = \mu e^{i\phi}$ the last two terms can be
also rewritten as $ 2\,\mu\,\cos(\phi+\delta)\,D$.  For the rest of
this section we will neglect the phases of $\mu$ and $\mathcal{D}$ and
assume that both quantities are real. Formally, one can implement this
condition in the scalar potential via a chiral rotation of the quark
fields in \eqn{eq:Ly}. This is described in detail
in~\cite{Fong:2013sba} (although it was later found that the solution
to the strong CP problem put forth in~\cite{Fong:2013sba} does not
hold~\cite{Fong:2013dnk}, the analysis of the relation between the
phase $\phi+\delta$ chiral rotations and the chiral anomaly remains
valid).  In \eqn{eq:V3} we require $\lambda>0$ in order to have a
potential bounded from below and $m^2 > 0$ to trigger SSB via $\langle
T\rangle \neq 0$.  From the r.h.s of \eqns{eq:Au}{eq:Du} we see that
both $A$ and $D$ are maximized for the symmetric configuration
$\langle \chi \rangle_s= \left(u_s,u_s,u_s\right)$. This would break
$\mathcal{G}_{LR}$ to the maximal subgroup $ H_{s}=SU(3)_{L+R}$.  The
minimum value $\langle D \rangle=0$ is obtained when one entry in
$\langle \chi \rangle$ vanishes, while $\langle A \rangle=0$ is
obtained when two entries vanish: $\langle \chi \rangle_h=
\left(0,0,u_h\right)$ (with $u_h = \mu/\sqrt{2\lambda}$). Therefore,
if $\lambda_A > 0 $ and a specific condition $\mu^2/m^2 <
F(\lambda,\lambda_A)$ is satisfied (where $F(\lambda,\lambda_A)$ is a
simple algebraic expression of its arguments, see~\cite{Nardi:2011st})
the global minimum is obtained for $\langle \chi \rangle_h$. This 
gives as little group the maximal $\mathcal{G}_\mathcal{F}$
subgroup $H_{h}=SU(2)_L\times SU(2)_R\times
U(1)$~\cite{CabibboMaiani:1970}.

\subsection{The one loop effective potential}
\label{sec:loop}

The fact that the tree level potential admits the vacuum configuration
$\langle \chi \rangle_h= \left(0,0,u_h\right)$, which is a good
approximation to the quarks Yukawa pattern, is encouraging, and one
could hope that some type of corrections could lift the two zeroes to
hierarchically suppressed entries. Clearly this would correspond to
breaking further the little groups of the tree level vacua. That is,
we need to understand if a stepwise breaking of the initial symmetry
group $\mathcal{G}_{LR}\to H_{h} \to$ nothing is possible.

In~\cite{Nardi:2011st} it was shown
that by adding to the tree level potential new terms of the form
$c_D\, \mu D\, \log D$ and $c_A\, A \log A$, with coefficients
$c_{D,A}< 1$, would do exactly this, lifting the two zeroes to
nonvanishing but exponentially suppressed entries. In the same paper
it was conjectured that the one loop corrected effective potential
might contain precisely terms of this type.

The non trivial task of computing the effective potential for the
scalar multiplet $Y$ was undertaken in~\cite{Espinosa:2012uu}.  We
first worked out the analytical expression of $V_{\rm eff}$, and next
we carried out a numerical study of its minima. We have found that the
vacuum structure of the tree level minimum remains stable against loop
corrections.  The reason for this stability can in fact be understood
on the basis of a theorem established by Georgi and Pais (GP) in the
late seventies~\cite{Georgi:1977hm}, stating that: {\it stepwise SSB
  can only occur via perturbative quantum corrections if there are
  non-Goldstone massless bosons in the tree approximation}. Since in
our case all massless modes correspond to Goldstone bosons, the little
groups $H_{s,h}$ cannot undergo further breaking because of loop
effects. (The Coleman-Weinberg model~\cite{Coleman:1973jx} is a well
known example of a tree level symmetry broken by loop effects.  This
can occur because classical scale invariance of the scalar potential
is assumed, and thus at the tree level all the scalars are massless
non-Goldstone modes. Therefore the occurrence of loop induced breaking
is in agreement with the GP theorem.)  In ref.~\cite{Espinosa:2012uu}
the GP theorem was extended to include also the case of perturbations
due to effective operators involving higher order invariants of $Y$,
and this lead to the conclusion that, in the absence of tree level
non-Goldstone massless modes: {\it no perturbative effect of any kind
  can further break the tree level little groups $H_{s,h}$ and lift
  the vanishing entries in $\langle\chi\rangle$}.

As a side remark, it should be noticed that the previous result does
not mean that one cannot write a {\it polynomial expression} in the
invariants which, upon minimization, yields a hierarchical pattern
$\langle \chi \rangle_{\rm exp}= \left(\epsilon',\epsilon,u\right)$
($\epsilon'\ll \epsilon\ll u$) and breaks $\mathcal{G}_\mathcal{F}$
completely.  This in fact can be done straightforwardly.  For a given
type of quarks, let us denote the experimental value of the invariants
as $T_{\rm exp}$, $A_{\rm exp}$, $D_{\rm exp}$. Then the nonnegative
polynomial $ P(Y)=\left(T- T_{\rm exp} \right)^2 +\frac{1}{\Lambda^4}
\left(A- A_{\rm exp} \right)^2 +\frac{1}{\Lambda^2}\left(D- D_{\rm
    exp} \right)^2$ is guaranteed to have its minimum in $\langle
\chi\rangle_{\rm exp}$~\cite{Alonso:2011yg}.  (Of course, with a
suitable set of independent invariants, this can be extended to the
full quark sector, so that all quark masses and mixing angles can be
trivially obtained via minimization of a suitable polynomial.)
However, expanding $P(Y)$ and truncating it to retain only the
renormalizable operators, one obtains that the $A$ invariant appears
with an overall negative sign $- 2 \left(A_{\rm exp}/\Lambda^4\right)
A$ and this implies that the minimum occurs for the symmetric solution
$\langle \chi \rangle_s= \left(u_s,u_s,u_s\right)$. Thus the
configuration $\langle \chi \rangle_{\rm exp}\sim
\left(\epsilon',\epsilon,u\right)$ is not obtained from a perturbation
of the hierarchical tree level vacuum $\langle \chi \rangle_h\sim
\left(0,0,u_t\right)$, meaning that the higher order terms dominate
over the renormalizable ones. Of course, this means that such an
approach would result in an incurable loss of predictivity.

\subsection{Symmetry breaking via  reducible representations}
\label{sec:reducible}

The results of the previous section make clear which way is left open
to get a phenomenologically viable pattern of vevs for the components
of the Yukawa field $Y$. Namely, the flavour symmetry
$\mathcal{G}_{LR}$ must be completely broken already at the tree
level.  For this, we need a non-minimal set of scalar fields in
reducible representations of the flavour group. In fact, breaking a
symmetry by means of reducible representations avoids at once the
issue of stability of the little groups that are maximal subgroup of
the original symmetry.  A minimal enlargement of the scalar sector
involves adding two multiplets, $Z_{L,R}$ transforming respectively
in the fundamental of one of the two group factors $SU(3)_L \times
SU(3)_R$, while being singlets under the other one:
\begin{equation}
  \label{eq:ZLZR}
 Z_L = (\mathbf{3},\mathbf{1}), \qquad Z_R = (\mathbf{1},\mathbf{3})\,.
\end{equation}
Let us write the most general $SU(3)_L \times SU(3)_R$ invariant potential 
involving $Z_L$, $Z_R$ and $Y=(\mathbf{3},\mathbf{\bar 3})$  
as:
\begin{equation}
\label{eq:VYZ}
V(Y,Z_L,Z_R) =V_{\cal I}+V_{\cal AR}+V_{\cal A}\,.
\end{equation}
$V_{\cal I}$ collects the so called {\it flavour irrelevant}
terms~\cite{Fong:2013dnk}. These are terms that are invariant under
accidental symmetries that are much larger than the flavour symmetry,
and thus the values of their vevs do not depend on any particular
flavour configuration.  For example $T$ \eqn{eq:Tu} is invariant under
$SO(18)$ which is broken down to $SO(17)$ after SFSB. A transformation
of this large subgroup can rotate $\langle \chi\rangle_s$ into
$\langle \chi\rangle_h$ leaving unchanged the ``length'' of the vev of
$Y$ (defined as $\sqrt{\langle T\rangle}$). Another example of flavour
irrelevant operators is $\left|Z_{L,R}\right|^2$ which carries a
$SO(6)$ symmetry.  All in all, the role of $V_{\cal I}$ is just that
of determining the ``length'' of the vevs for $Y,Z_L,Z_R$, without
contributing to the determination of any specific flavour
direction. Therefore we will omit writing here its explicit form,
that can be found  in~\cite{Espinosa:2012uu}.

Those terms in the potential that tend to break the symmetry to the
largest maximal little group (in this case $SU(3)_{L+R}$ with eight
generators) are defined as {\it attractive} ($\cal A$), while terms
that tend to break the symmetry to the smallest maximal little group
(in this case $SU(2)_L\times SU(2)_R\times U(1)$ with seven
generators) are called {\it repulsive} ($\cal R$).  Hermitian
monomials can be attractive or repulsive depending if their (real)
couplings are negative or positive, and are included in  $V_{\cal AR}$:
\begin{equation}
  V_{\cal AR} = \lambda_AA + g_{R}   |Y Z_R|^2 + 
g_L \, |Y^{\dagger} Z_L|^2\,,    
\end{equation}
where we have adopted for the modulus square notation the convention
$\left|X\right|^2= X^\dagger X$. For example (see
section~\ref{sec:tree}) the sign of $\lambda_A$ determines if the $A$
invariant is attractive or repulsive.  Operators which correspond to
non-Hermitian monomials are included in $V_{\cal A}$:
\begin{eqnarray}
  \label{eq:VA}
V_{\cal A}&=& \tilde\mu \mathcal{D} + \tilde\nu Z^\dagger_L Y Z_R 
+ {\rm H.c.}\\
\nonumber
&=&  2\,\mu\,D\, \cos \delta +
 2\,\nu\, \left|Z_{L}^{\dagger}YZ_{R}\right| \cos\phi_{LR}\,,
\end{eqnarray}
where in the second line $\mu = |\tilde\mu|$ and $\nu = |\tilde \nu|$.
Non-Hermitian monomials are always attractive, as is the case for
${\cal D}$: when the vev $\langle {\cal D}\rangle$ is nonvanishing,
minimization drives its phase $\delta\to \pi$
($\cos\delta= -1$). Then the minimum gets lowered for the
largest possible value of $D$, that is obtained for $\langle
\chi\rangle_s$, corresponding to the largest little group $H_s=SU(3)_{L+R}$.
Let us now define 
\begin{eqnarray}
  \label{eq:newvevs}
% \nonumber
\langle Y\rangle &=& 
v_Y\,{\rm diag}\left(\epsilon',\epsilon,y\right), \\
\langle Z_{L}\rangle &=& v_L\,
\left(z_{L},\epsilon'_{L},\epsilon_L\right)\\
\langle Z_{R}\rangle &=& v_R\,
\left(z_{R},\epsilon'_{R},\epsilon_R\right)
\end{eqnarray}
with $\epsilon^2+{\epsilon'}^2+ y^2=1$ and analogously for the entries
in $\langle Z_{L,R}\rangle$. Let us see if a hierarchical solution
$\epsilon,\epsilon'\ll y$, $\epsilon_{L,R},\epsilon'_{L,R}\ll
z_{L,R}$, can be obtained.  $V_{\cal I}$ in \eqn{eq:VYZ} fixes the
`lengths' $v_Y$ and $v_{L,R}$, and vanishes, so we need to consider
only the effect of $V_\epsilon\equiv V_{\cal AR}+V_{\cal A}$.  For
$\lambda_A,g_{L,R}> 0 $ $V_{\cal AR}$ is always positive, and thus it
is minimized when it vanishes, which occurs when the vevs of $Z_{L,R}$
are misaligned with respect to the vev of $Y$, as for example 
$\langle Y\rangle = v_Y\,{\rm diag}\left(0,0,1\right)$, $\langle
Z_{L}\rangle =v_L\, \left(c_{L},s_{L},0\right)$ and $\langle
Z_{R}\rangle =v_R\,
\left(c_{R},s_{R},0\right)$  ($c_{L,R}^2+s_{L,R}^2=1$).  However, such a
configuration would also imply $V_{\cal A}=0$, while configurations
yielding $V_{\cal A}<0$ would result in a lower minimum.  We then
learn that the parameters $\mu$ and $\nu$ appearing in $V_{\cal A}$
can play a crucial role in lifting the vanishing entries.  A simple
illustrative example of their action can be given by setting for
simplicity $v_{Y}=v_{L}=v_{R}$ and $\lambda_A=g_{L}=g_{R}$.  Solving
for the extremal conditions $\partial V_{\epsilon}/\partial\epsilon
=\partial V_{\epsilon}/\partial\epsilon' =\partial
V_{\epsilon}/\partial\epsilon_{L,R} =\partial
V_{\epsilon}/\partial\epsilon_{L,R}'=0$, and truncating to terms
$\mathcal{O}\left(\epsilon^{2}\right)$ we obtain a unique solution for
the global minimum:
\begin{equation}
\epsilon =  \frac{\lambda_A\, \nu\, v_Y}{3\lambda_A^2\,v_Y^2-\mu^{2}}\,, 
\qquad 
\epsilon' =  \frac{\mu}{\lambda_A\, v_Y}\, \epsilon\,. 
\label{eq:finalvevs}
\end{equation}
At this order, the other parameters vanish ($\epsilon_{L,R} =
\epsilon_{L,R}'=0$), and the potential minimum is $V_\epsilon^{{\rm
    min}}=- \nu\, v_Y^3\,\epsilon$.  \Eqn{eq:finalvevs} shows that a
hierarchy $\epsilon' \sim 10^{-2}\cdot \epsilon \sim 10^{-4}$, which
would fit well the observed values of  $\langle \chiu\rangle$,  can be
obtained by taking $\nu \sim \mu \sim 10^{-2}\cdot \lambda_A\, v_Y$.

\section{Two Yukawa fields: generating quark mixing}
\label{sec:mixing}

In this section we extend the study of SFSB by including both the $u$
and $d$ sectors. The flavour symmetry is then the full flavour group
${\cal G}_{\cal F}$~\eqn{eq:GF}, under which the $u$- and $d$-type
Yukawa multiplets transform as:
$Y_{u}\sim\left(\mathbf{3},\mathbf{\bar 3},\mathbf{1}\right)$ and
$Y_{d}\sim\left(\mathbf{3},\mathbf{1},\mathbf{\bar 3}\right)$.

\subsection{Minimization of the potential for $Y_u$ and $Y_d$}
\label{sec:mixingYY}

With just the two multiplets $Y_u$ and $Y_d$ we can write only one
invariant that is not flavour irrelevant and that couples the two Yukawa fields 
(for the complete expression for $V(Y_u,Y_d)$ see~\cite{Nardi:2011st}):
\begin{equation} 
   \label{eq:Tud}
T_{ud}  = {\rm Tr}( Y_uY_u^\dagger Y_dY_d^\dagger )\ =\ 
 {\rm Tr}\left(K^\dagger \chiusq  K  \chidsq \right) \,,  
\end{equation}
where $K$ is a unitary matrix of fields that in terms of the singular
value decomposition parameterization \eqn{eq:biunitary} is given by
$K={\cal V}_u {\cal V}_d^\dagger$.  The vev of $K$ describes the
mismatch between the two basis in which $\langle Y_{u}\rangle$ and
$\langle Y_{d}\rangle$ are diagonal, and after ordering $\langle
\chiu\rangle $ and $\langle \chid\rangle $ in the same way (e.g. with
increasing size of their entries) we can make the identification
$\langle K \rangle = V_{CKM}$.  Being a Hermitian monomial, $T_{ud}$
can be attractive or repulsive depending on the sign of its coupling
$\lambda_{ud}$ (respectively negative or positive).  At fixed lengths
($ \sqrt{\langle T_u\rangle}, \sqrt{\langle T_d\rangle} ={\rm
  const.}$) when $\langle Y_u\rangle $ and $\langle Y_d\rangle $ are
``aligned'' (i.e. they are diagonal in the same basis and with the
same ordering) the value of $\langle T_{ud}\rangle $ is maximum, while
it is minimum when in a given basis they are ``anti-aligned''
(diagonal but with opposite ordering).  It is then clear that there
are only two options to extremize $T_{ud}$: if its coupling
$\lambda_{ud}$ is negative, alignment is selected and we obtain
$\langle K \rangle = I$. $\lambda_{ud}>0$ selects instead the
anti-aligned configuration, which means that $\langle K \rangle$ is
anti-diagonal.  The first possibility can be considered to give a
reasonable first approximation to $V_{CKM}$. However, also in this
case there is no type of perturbative correction that can produce
departures from an exact $\langle K \rangle = I$ and generate small
mixing angles. A simple intuitive way to understand this is to notice
that with only two ``directions'' in flavour space ($Y_{u}$ and
$Y_{d}$) there is just a single relative ``angle'', which gets fixed
by minimization of the potential.  This implies that alignment or
anti-alignment are the only possibilities for extremization.

One additional remark is in order.  With a negative coupling
($\lambda_{ud} <0$) the term in \eqn{eq:Tud} besides aligning $\langle
Y_{u}\rangle $ and $\langle Y_{d}\rangle $ has also another important
effect: the naive hierarchical pattern $y\sim 1$, $\epsilon \sim
\frac{\nu}{v_Y}$, $\epsilon'\sim \frac{\mu}{v_Y}\epsilon$
that we have 
derived at the end of section~\ref{sec:reducible}, see \eqn{eq:finalvevs}, 
gets amplified by the effect of $T_{ud}$.  Numerically, we find that instead 
in the presence of  $T_{ud}$ and with $\lambda_{ud} <0$, 
$\frac{\mu}{v} \sim 10^{-1} |\lambda_{ud}|$ is enough to generate
sufficiently strong Yukawa hierarchies as the ones observed.  For an 
explanation of how this dynamical enhancement comes about we refer
to~\cite{Fong:2013dnk}.

\subsection{Quark mixing via  reducible representations}
\label{sec:mixreducible} 

As we have seen in section \ref{sec:reducible}, adding new scalar
representations is in any case necessary to generate hierarchical
Yukawa matrices with nonvanishing entries.  We will now argue that
this can also cure the troublesome result $V_{CKM}=I$.  We have seen
that the pair of scalar multiplets $Z_{L,R}$ transforming respectively
in the fundamental of the $SU(3)_{L,R}$ factors of ${\cal G}_{LR}$
\eqn{eq:GLR} suffices to generate hierarchical entries for one Yukawa
multiplet. Then, as a first attempt to generate a vev $\langle K
\rangle$ with a non trivial structure, we can try to introduce, as a
minimal number of additional fields, the following three multiplets:
$Z_{Q_1}\sim\left(\mathbf{3},\mathbf{1},\mathbf{1}\right)$,
$Z_{u}\sim\left(\mathbf{1},\mathbf{3},\mathbf{1}\right)$ and
$Z_{d}\sim\left(\mathbf{1},\mathbf{1},\mathbf{3}\right)$, where in
parenthesis we have given the transformation properties under
$\mathcal{G}_\mathcal{F}$.  Therefore we have in total three `vectors'
$Z_{Q_1},Y_u$ and $Y_d$ transforming under the L-handed factor
$SU(3)_Q$ of ${\cal G}_{\cal F}$, while $Z_u$ and $Z_d$ transform
respectively in the fundamental of the two R-handed factors.  The
study of the most general tree level potential for these five
multiplets in terms of attractive/repulsive operators is a bit
cumbersome, and we refer to section IV of ref.~\cite{Fong:2013dnk} for
details. The result is that while this minimal set suffices for
obtaining hierarchical solutions for both $\langle \chi_u\rangle$ and
$\langle \chi_d\rangle$, only one nontrivial mixing angle
($\theta_{23}$) is generated, which also implies that there are no
sources of CP violation in the ground state for the mixing matrix
$\langle K\rangle$.

\subsection{Mass hierarchies, CKM mixings and  CP violation}
\label{sec:CP}

In order to generate the other two nonvanishing mixing angles
($\theta_{12}$ and $\theta_{13}$) at least one additional multiplet
transforming in the fundamental of $SU(3)_Q$ is needed: $Z_{Q_2}\sim
\left(\mathbf{3},\mathbf{1}, \mathbf{1}\right)$.  This can be
intuitively understood by observing that four independent
``directions'' in L-handed flavour space ($Z_{Q_{1}},Z_{Q_{2}}\, Z_u$,
$Z_d$) constitute the minimum number required to define three relative
``angles''. While the expression for the full $\mathcal{G}_{\cal F}$
invariant potential becomes a bit involved, a large number of terms
are flavour irrelevant, and understanding the dynamical action of the
relevant invariants is still a manageable task. This task has been
carried out in~\cite{Fong:2013dnk}. Quite interestingly, we have found
that the same field content that ensures three nonvanishing mixing
angles, is also sufficient to ensure that at the potential minimum the
vev $\langle K \rangle \sim V_{CKM}$ contains one CP violating
phase. Let us stress, for the seek of clarity, that in the present
case the fact that the potential minimum is CP violating has nothing
to do with the notion of {\it spontaneous CP violation}. In fact, in
our scenario the scalar potential violates CP from the start because
of the presence of a certain number of physical complex phases (four,
to be exact).  The important point thus is that at the minimum there
is no restoration of CP symmetry, something that could have happened
if, by some accident, all the field vevs had flown towards real
values.

\subsection{A numerical example}
\label{numerical}

The final verdict if spontaneous breaking of the flavour symmetry is
able to account for the entire set of observables in the quark sector
can eventually be settled only by means of numerical minimization of
the full ${\cal G}_{\cal F}$-invariant scalar potential. This issue
has been addressed in~\cite{Fong:2013dnk}. However, no attempts to
perform multidimensional global fits to the SM observables where
carried out, something that would have required a prohibitive amount
of CPU time.  The simpler approach followed in~\cite{Fong:2013dnk} was
that of assuming a simple set of values for most of the flavour
irrelevant parameters and then, by varying the remaining (crucial)
ones, attempting to approximate the experimental values of the
observables of the quark sector.  Of course, carrying out successfully
this procedure has been rendered feasible by a good understanding of
the role of each term in the potential, an understanding that we have
gained by inspection of several partial analytical results.  An
example of the type of results that can be obtained is given below.
Since we have not reproduced here the (rather lengthy) expression for
the most general ${\cal G}_{\cal F}$ invariant potential for the $Y$
and $Z$ fields, we also do not recopy the numerical values of the
input parameters.  Since it is a quite relevant point we mention,
however, that all parameters have been taken to be ${\cal O}(1)$, with
no hierarchies among them larger than ${\cal O}(10^{-1})$.  We have
found input values satisfying the above conditions for which the
resulting parameters of the SM quark sector are
\begin{eqnarray}
\langle Y_{u}\rangle 
& = & v_{u}\>{\rm diag}\left(0.0003,0.009,1.4\right),
\nonumber \\
\langle Y_{d}\rangle 
& = & v_{d}\>{\rm diag}\left(0.0007,0.02,1.2\right),
\nonumber \\
|\langle K\rangle|  & = & \left(\begin{array}{ccc}
0.974 & 0.223 & 0.027\\
0.224 & 0.974 & 0.042\\
0.017 & 0.046 & 0.999
\end{array}\right),\nonumber \\
J & = & 2.9\times10^{-5}, 
\label{eq:output1}
\end{eqnarray}
where $J$ in the last line is the Jarskog invariant~\cite{Jarlskog:1985ht}.
Let us note that having the largest entries in $\langle Y_{u,d}\rangle$ of
similar size, which follows from the fact that we have set $v_{d}=v_{u}$
from the start, does not
constitute any real problem. The value of the $b$-quark mass can be easily
suppressed by means of a $U(1)$ spurion vev, along the lines described
for example in \cite{Nardi:2011st}, or by extending the Higgs sector
to a two doublets model with $\langle H_d\rangle \ll \langle
H_u\rangle$.

\section{Conclusions}
\label{sec:conclusions}

In this contribution we have reported the attempt developed in the
three papers~\cite{Nardi:2011st,Espinosa:2012uu,Fong:2013dnk} to
explain the values of the parameters of the SM quark sector (four mass
ratios, three mixing angles and one CP violating phase) starting from
the idea that the complete breaking of the quark flavour symmetry can
result as the dynamical effect of driving a suitable scalar potential
towards its minimum. We have identified the minimum set of multiplets
in simple (fundamental and bifundamental) representations of the group
needed to break ${\cal G}_{\cal F}\to 0$, and we have shown that this
same set of fields is also sufficient to generate one weak CP
violating phase. Besides the quantitative results, through this study
we have gained important qualitative understandings of various
mechanisms that might underlie some of the most puzzling features of
the SM quark sector. We list them in
what we think is their order of importance. \\  [-0.2cm]

1. $K = V_{CKM} \approx I$.\ The interaction between the two Yukawa
fields $Y_u$ and $Y_d$ tends to generate an exact alignment of their
vevs in flavour space, resulting in $V_{CKM} = I$
~\cite{Anselm:1996jm}.  If the interaction is repulsive ($\lambda_{ud}
> 0$) the alignment occurs when the eigenvalues of the two matrices
are ordered by size in an opposite way.  When the interaction is
attractive ($\lambda_{ud} < 0$) the alignment occurs when the ordering
is the same. This second possibility is the one observed in nature. To
generate three nonvanishing mixing angles, that is to (slightly)
misalign $Y_u$ and $Y_d$ in all flavour directions, at least two other
multiplets transforming under the L-handed factor $SU(3)_Q$ are
needed.  Their presence will induce perturbation in the exact
alignment, but if the $Y_u$-$Y_d$ interaction is sufficiently strong,
an approximate
form $V_{CKM} \approx I$ will be maintained.\\  [-0.2cm]

2. {\it Yukawa hierarchies.}\ Hierarchies between the different
entries in $\langle Y_u\rangle $ and $\langle Y_d\rangle $ are seeded
by taking for a subset of the dimensional parameters values somewhat
smaller than the overall scale of the vevs: $\mu, \nu \sim
v_Y/10$. This can be justified by the fact that when these parameters
are set to zero, the scalar potential gains some additional $U(1)$
symmetries.  The initial (mild) suppression of some entries in
$\langle Y_{u,d}\rangle$ can get amplified down to the observed values
of the quark mass ratios by dynamical effects.  Hierarchical Yukawa
couplings can then be generated without
strong hierarchies in the fundamental parameters.  \\  [-0.2cm]

3. {\it CP violation.}\ Once the flavour symmetry is completely
broken, generating the CKM CP violating phase does not require
complicating further the model.  The set of scalar multiplets needed
to obtain ${\cal G}_{\cal F}\to 0 $ ensures that several complex
phases cannot be removed regardless of field redefinitions, and this
ensures that the scalar potential contains CP violating terms.  For
generic values of these phases, a CP violating
ground state for $V(Y_q,Z_q,Z_{Q_{1,2}})$ is obtained.  \\  [-0.2cm]

Indeed, one could object that in our construction there are many more
fundamental parameters than there are observables. This of course
affects its predictivity, and in some respects also its elegance.  We
cannot object to such a criticism, but it is worth stressing that the
proliferation of parameters is a result of the usual quantum field
theory prescription for building renormalizable Lagrangians: we have
identified the minimum number of multiplets needed to break completely
${\cal G}_{\cal F}$, and next we have simply written down the complete
set of renormalizable operators allowed by the symmetry. After all, as
it has been argued e.g.  in~\cite{Duque:2008ah}, the apparent lack of
simple relations between the observables of the quark sector might
well be due to the fact that, as in our case, they are determined by a
very large number of fundamental parameters.

Direct evidences of the scenario we have been studying might arise
from the fact that if the flavour symmetry is global, then spontaneous
symmetry breaking implies the presence of Nambu-Goldstone bosons that
could show up in yet unseen hadron decays or in rare flavour violating
processes. If the flavour symmetry is instead gauged, then to ensure
the absence of gauge anomalies additional fermions must be
introduced~\cite{Grinstein:2010ve}, and their detection could then
represent a smoking gun for this type of models. All this remains,
however, a bit speculative, especially because the theory provides no
hint of the scale at which the flavour symmetry gets broken, and very
large scales would suppress most, if not all, types of signatures.

\section*{Acknowledgments}
It is a pleasure to thank my colleagues Jose Ramon Espinosa and Chee
Sheng Fong for their precious collaboration in developing this
research project.
% writing ref.~\cite{Espinosa:2012uu}. 
% C. F. S.  is supported by Funda\c{c}\~ao de Amparo \`a Pesquisa do Estado de
% S\~ao Paulo (FAPESP).
This work  is supported in part
by the research grant % "Theoretical Astroparticle Physics" 
number 2012CPPYP7 under the program PRIN 2012 funded by the Italian
``Ministero dell’'  Istruzione, Universit\'a e della Ricerca'' (MIUR) and
by the INFN ``Iniziativa Specifica'' 
% Theoretical Astroparticle Physics  
TAsP-LNF.

 
 
%% References 
 
%   \nocite{*} 
% \bibliographystyle{elsarticle-num} 

% \bibliography{SFSB} 

\begin{thebibliography}{10}

\bibitem{Froggatt:1978nt}
C.~Froggatt and H.~B. Nielsen, ``{Hierarchy of Quark Masses, Cabibbo Angles and
  CP Violation},''
\href{http://dx.doi.org/10.1016/0550-3213(79)90316-X}{{\em Nucl.Phys.}
  {\bfseries B147} (1979) 277}.    \\[-20pt]
%%CITATION = NUPHA,B147,277;%%.



\bibitem{Chivukula:1987py}
R.~S. Chivukula and H.~Georgi, ``{Composite Technicolor Standard Model},''
\href{http://dx.doi.org/10.1016/0370-2693(87)90713-1}{{\em Phys.Lett.}
  {\bfseries B188} (1987) 99}. \\[-25pt]
%%CITATION = PHLTA,B188,99;%%.

\bibitem{Nardi:2011st}
E.~Nardi, ``{Naturally large Yukawa hierarchies},''
  \href{http://dx.doi.org/10.1103/PhysRevD.84.036008}{{\em Phys.Rev.}
  {\bfseries D84} (2011) 036008},
\href{http://arxiv.org/abs/1105.1770}{{\ttfamily arXiv:1105.1770 [hep-ph]}}. \\[-25pt]
%%CITATION = ARXIV:1105.1770;%%.

\bibitem{Espinosa:2012uu}
J.~R. Espinosa, C.~S. Fong, and E.~Nardi, ``{Yukawa hierarchies from
  spontaneous breaking of the $SU(3)_L\times SU(3)_R$ flavour symmetry?},''
  \href{http://dx.doi.org/10.1007/JHEP02(2013)137}{{\em JHEP} {\bfseries 1302}
  (2013) 137},
\href{http://arxiv.org/abs/1211.6428}{{\ttfamily arXiv:1211.6428 [hep-ph]}}. \\[-25pt]
%%CITATION = ARXIV:1211.6428;%%.

\bibitem{Fong:2013dnk}
C.~S. Fong and E.~Nardi, ``{Quark masses, mixings, and CP violation from
  spontaneous breaking of flavor $SU(3)^{3}$},''
  \href{http://dx.doi.org/10.1103/PhysRevD.89.036008}{{\em Phys.Rev.}
  {\bfseries D89} no.~3, (2014) 036008},
\href{http://arxiv.org/abs/1307.4412}{{\ttfamily arXiv:1307.4412 [hep-ph]}}. \\[-25pt]
%%CITATION = ARXIV:1307.4412;%%.

\bibitem{Michel:1971th}
L.~Michel and L.~Radicati, ``{Properties of the breaking of hadronic internal
  symmetry},''
\href{http://dx.doi.org/10.1016/0003-4916(71)90079-0}{{\em Annals Phys.}
  {\bfseries 66} (1971) 758--783}. \\[-25pt]
%%CITATION = APNYA,66,758;%%.

\bibitem{CabibboMaiani:1970}
N.~Cabibbo and L.~Maiani, ``{in {\it Evolution of particle physics}},''
{\em Academic Press} {\bfseries 50 App. I} (1970) 68--72. \\[-25pt]
%%CITATION = APNYA,66,758;%%.

\bibitem{Anselm:1996jm}
A.~Anselm and Z.~Berezhiani, ``{Weak mixing angles as dynamical degrees of
  freedom},'' \href{http://dx.doi.org/10.1016/S0550-3213(96)00597-4}{{\em
  Nucl.Phys.} {\bfseries B484} (1997) 97--123},
\href{http://arxiv.org/abs/hep-ph/9605400}{{\ttfamily arXiv:hep-ph/9605400
  [hep-ph]}}. \\[-25pt]
%%CITATION = HEP-PH/9605400;%%.

\bibitem{Berezhiani:2001mh}
Z.~Berezhiani and A.~Rossi, ``{Flavor structure, flavor symmetry and
  supersymmetry},'' \href{http://dx.doi.org/10.1016/S0920-5632(01)01527-4}{{\em
  Nucl.Phys.Proc.Suppl.} {\bfseries 101} (2001) 410--420},
\href{http://arxiv.org/abs/hep-ph/0107054}{{\ttfamily arXiv:hep-ph/0107054
  [hep-ph]}}. \\[-25pt]
%%CITATION = HEP-PH/0107054;%%.

\bibitem{Koide:2008qm}
Y.~Koide, ``{Phenomenological Meaning of a Neutrino Mass Matrix Related to
  Up-Quark Masses},'' \href{http://dx.doi.org/10.1103/PhysRevD.78.093006}{{\em
  Phys.Rev.} {\bfseries D78} (2008) 093006},
\href{http://arxiv.org/abs/0809.2449}{{\ttfamily arXiv:0809.2449 [hep-ph]}}. \\[-25pt]
%%CITATION = ARXIV:0809.2449;%%.

\bibitem{Koide:2008tr}
Y.~Koide, ``{Charged Lepton Mass Relations in a Supersymmetric Yukawaon
  Model},'' \href{http://dx.doi.org/10.1103/PhysRevD.79.033009}{{\em Phys.Rev.}
  {\bfseries D79} (2009) 033009},
\href{http://arxiv.org/abs/0811.3470}{{\ttfamily arXiv:0811.3470 [hep-ph]}}. \\[-25pt]
%%CITATION = ARXIV:0811.3470;%%.

\bibitem{Koide:2012fw}
Y.~Koide and H.~Nishiura, ``{Yukawaon Model with U(3)$\times$S$_3$ Family
  Symmetries},'' \href{http://dx.doi.org/10.1016/j.physletb.2012.05.014}{{\em
  Phys.Lett.} {\bfseries B712} (2012) 396--400},
\href{http://arxiv.org/abs/1202.5815}{{\ttfamily arXiv:1202.5815 [hep-ph]}}. \\[-25pt]
%%CITATION = ARXIV:1202.5815;%%.

\bibitem{Feldmann:2009dc}
T.~Feldmann, M.~Jung, and T.~Mannel, ``{Sequential Flavour Symmetry
  Breaking},'' \href{http://dx.doi.org/10.1103/PhysRevD.80.033003}{{\em
  Phys.Rev.} {\bfseries D80} (2009) 033003},
\href{http://arxiv.org/abs/0906.1523}{{\ttfamily arXiv:0906.1523 [hep-ph]}}. \\[-25pt]
%%CITATION = ARXIV:0906.1523;%%.

\bibitem{Albrecht:2010xh}
M.~Albrecht, T.~Feldmann, and T.~Mannel, ``{Goldstone Bosons in Effective
  Theories with Spontaneously Broken Flavour Symmetry},''
  \href{http://dx.doi.org/10.1007/JHEP10(2010)089}{{\em JHEP} {\bfseries 1010}
  (2010) 089},
\href{http://arxiv.org/abs/1002.4798}{{\ttfamily arXiv:1002.4798 [hep-ph]}}. \\[-25pt]
%%CITATION = ARXIV:1002.4798;%%.

\bibitem{Grinstein:2010ve}
B.~Grinstein, M.~Redi, and G.~Villadoro, ``{Low Scale Flavor Gauge
  Symmetries},'' \href{http://dx.doi.org/10.1007/JHEP11(2010)067}{{\em JHEP}
  {\bfseries 1011} (2010) 067},
\href{http://arxiv.org/abs/1009.2049}{{\ttfamily arXiv:1009.2049 [hep-ph]}}. \\[-25pt]
%%CITATION = ARXIV:1009.2049;%%.

\bibitem{Alonso:2011yg}
R.~Alonso, M.~Gavela, L.~Merlo, and S.~Rigolin, ``{On the scalar potential of
  minimal flavour violation},''
  \href{http://dx.doi.org/10.1007/JHEP07(2011)012}{{\em JHEP} {\bfseries 1107}
  (2011) 012},
\href{http://arxiv.org/abs/1103.2915}{{\ttfamily arXiv:1103.2915 [hep-ph]}}. \\[-25pt]
%%CITATION = ARXIV:1103.2915;%%.

\bibitem{deMedeirosVarzielas:2011wx}
I.~de~Medeiros~Varzielas, ``{Non-Abelian family symmetries in Pati-Salam
  unification},'' \href{http://dx.doi.org/10.1007/JHEP01(2012)097}{{\em JHEP}
  {\bfseries 1201} (2012) 097},
\href{http://arxiv.org/abs/1111.3952}{{\ttfamily arXiv:1111.3952 [hep-ph]}}. \\[-25pt]
%%CITATION = ARXIV:1111.3952;%%.

\bibitem{Mohapatra:2012km}
R.~N. Mohapatra, ``{Gauged Flavor, Supersymmetry and Grand Unification},''
  \href{http://dx.doi.org/10.1063/1.4742073}{{\em AIP Conf.Proc.} {\bfseries
  1467} (2012) 7--14},
\href{http://arxiv.org/abs/1205.6190}{{\ttfamily arXiv:1205.6190 [hep-ph]}}. \\[-25pt]
%%CITATION = ARXIV:1205.6190;%%.

\bibitem{Alonso:2013nca}
R.~Alonso, M.~Gavela, G.~Isidori, and L.~Maiani, ``{Neutrino Mixing and Masses
  from a Minimum Principle},''
  \href{http://dx.doi.org/10.1007/JHEP11(2013)187}{{\em JHEP} {\bfseries 1311}
  (2013) 187},
\href{http://arxiv.org/abs/1306.5927}{{\ttfamily arXiv:1306.5927 [hep-ph]}}. \\[-25pt]
%%CITATION = ARXIV:1306.5927;%%.

\bibitem{Fong:2013sba}
C.~S. Fong and E.~Nardi, ``{Spontaneous Breaking of Flavor Symmetry Avoids the
  Strong CP Problem},''
  \href{http://dx.doi.org/10.1103/PhysRevLett.111.061601}{{\em Phys.Rev.Lett.}
  {\bfseries 111} no.~6, (2013) 061601},
\href{http://arxiv.org/abs/1305.1627}{{\ttfamily arXiv:1305.1627 [hep-ph]}}. \\[-25pt]
%%CITATION = ARXIV:1305.1627;%%.

\bibitem{Georgi:1977hm}
H.~Georgi and A.~Pais, ``{Natural Stepwise Breaking of Gauge and Discrete
  Symmetries},''
\href{http://dx.doi.org/10.1103/PhysRevD.16.3520}{{\em Phys.Rev.} {\bfseries
  D16} (1977) 3520}. \\[-25pt]
%%CITATION = PHRVA,D16,3520;%%.

\bibitem{Coleman:1973jx}
S.~R. Coleman and E.~J. Weinberg, ``{Radiative Corrections as the Origin of
  Spontaneous Symmetry Breaking},''
\href{http://dx.doi.org/10.1103/PhysRevD.7.1888}{{\em Phys.Rev.} {\bfseries D7}
  (1973) 1888--1910}. \\[-25pt]
%%CITATION = PHRVA,D7,1888;%%.

\bibitem{Jarlskog:1985ht}
C.~Jarlskog, ``{Commutator of the Quark Mass Matrices in the Standard
  Electroweak Model and a Measure of Maximal CP Violation},''
\href{http://dx.doi.org/10.1103/PhysRevLett.55.1039}{{\em Phys.Rev.Lett.}
  {\bfseries 55} (1985) 1039}. \\[-25pt]
%%CITATION = PRLTA,55,1039;%%.

\bibitem{Duque:2008ah}
L.~F. Duque, D.~A. Gutierrez, E.~Nardi, and J.~Norena, ``{Fermion mass
  hierarchy and non-hierarchical mass ratios in SU(5) x U(1)(F)},''
  \href{http://dx.doi.org/10.1103/PhysRevD.78.035003}{{\em Phys.Rev.}
  {\bfseries D78} (2008) 035003},
\href{http://arxiv.org/abs/0804.2865}{{\ttfamily arXiv:0804.2865 [hep-ph]}}. \\[-25pt]
%%CITATION = ARXIV:0804.2865;%%.

\end{thebibliography}
% \bibliographystyle{utphys} 
% \end{document} 

\providecommand{\href}[2]{#2}\begingroup\raggedright
\endgroup
 
%% The provided style file elsarticle-num.bst formats references 
%% in the required Procedia style 
 
 
 
 
 
% \end{thebibliography} 
 
\end{document}